\documentstyle[epsf,aps]{revtex}

\def\dslash{\partial\!\!\!/}
\begin{document}

\preprint{\vbox {\hspace*{\fill} DOE/ER/40762-141\\
          \hspace*{\fill} U. of MD PP \98-076}} 
\vspace{.5in}

\title {Quantum Number Exotic Hybrid Mesons\\ and Large \boldmath {$N_c$} QCD}

\author{Thomas D. Cohen}

\address{Department of 
Physics, University of~Maryland, College~Park, MD~20742-4111}

\date{U. of MD PP\#98-076~~~DOE/ER/40762-141~~~January 1998}

\maketitle

\vspace{.25in}

\begin{abstract}
Exotic  mesons with quantum numbers of 
a hybrid ($\overline{q} q g$) such as $J^{PC} =
1^{-+}$are shown to exist as narrow resonant states in the  large $N_c$ 
limit of QCD.  The width of these states is proportional to  $1/N_c$.  
Possible phenomenological implications for the world of $N_c=3$ are discussed.
\end{abstract}
 
\vspace{.25in}
 
There has been considerable recent theoretical interest in hybrid mesons
both from the perspective of models\cite{m1,m2,m3,m4,m5,m6,m7,m8} and lattice
calculations\cite{l1,l2,l3,l4,l5,l6}.  In the language of the quark model, such mesons are composed of 
constituent quark and antiquark and one valence gluon.   Such states are of
interest in large measure because they depend on physics beyond the simplest
quark model.  
Ideally one would like to
understand such a state in terms of the underlying fundamental theory of strong
interactions---QCD.  However, this is problematic.  Indeed,
the precise meaning of constituent quarks and valence gluons in terms of QCD is
obscure. It is not clear just how the current quarks and gluons of QCD form into
effective constituent quarks and
gluons.  Moreover, from the experimental side there is generally ambiguity
in attempting to distinguish a hybrid from  an ordinary meson.  This has led 
to a focus on so-called quantum number exotic hybrid states.  Such states have
quantum numbers   which cannot be constructed from a
single quark and antiquark; for example,  $J^{P C} = 1^{- +}$ or $0^{- +}$. As such they are 
automatically exotic states from the
perspective of the quark model.  
 Recently, the E852 Collaboration at Brookhaven
has  reported evidence of the existence of a $J^{P C} = 1^{- +}$ resonance.
The reported mass is 1370 MeV  and the width 385 MeV\cite{e1}.  

>From the theoretical side, it is 
clear that one can construct states in QCD which have the quantum numbers 
$J^{P C} = 1^{- +}$ or $0^{- +}$.  However, this does not necessarily imply 
the existence of exotic hadrons.  After all, a hadron is not just any
state of the theory---it is   a resonant state and 
a relatively narrow one.  To be unambiguously identified as a hadron
the state must live for a considerable
period (on hadronic time scales) before decaying.  Thus the question of
whether QCD has quantum number exotic hybrid mesons ultimately comes down to 
whether there exist long-lived resonant states with exotic
quantum numbers. 

An important theoretical issue
is whether one can deduce the existence of narrow
exotic states directly from  QCD.   This problem  is qualitatively similar to 
trying to infer the existence of ordinary non-exotic mesons as narrow states
directly from QCD.   In the case of ordinary mesons, however, one can study QCD 
in the limit when the number of colors
($N_c$) becomes large,  and use  $1/N_c$ expansion techniques to deduce
 meson properties.  

The  $1/N_c$ expansion was introduced in QCD by
 `t Hooft as a way to formulate a systematic expansion  for low energy
 QCD \cite{NC1}. Properties of mesons  in large $N_c$ are reviewed  nicely
 in references \cite{NC2,NC3}.  Using large $N_c$ QCD one infers   
the existence of narrow mesons \cite{NC2,NC3} (at least in a large $N_c$ world.) 
 The essential premise is that the hypothetical
large $N_c$ world is qualitatively similar to the $N_c = 3$ world; one
can systematically improve the description by working to higher order in
$1/N_c$.  Ordinary mesons
 at large $N_c$
have  widths which are proportional to $1/N_c$ \cite{NC2,NC3}.
 To the extent that the large $N_c$ approximation is valid for the real world this
explains the existence of narrow mesonic states.

This raises an obvious question:  How do states with the quantum numbers 
of exotic hybrid mesons behave in the large $N_c$ limit?  The central purpose of the
present letter is to show that, as in the case of non-exotic mesons, exotic
hybrid mesons
exist and 
are narrow in the large $N_c$ limit---with a width  $ \sim 1/N_c$.  Indeed, 
it will be  shown here that from the perspective of large $N_c$ QCD there is 
nothing particularly exotic about quantum number exotic hybrid mesons.  
They have {\it
all} of the same $N_c$ dependences as ordinary mesons:  the vertex coupling 
$n_h$ quantum number exotic hybrids and $n_m$ ordinary mesons scale as
\begin{equation}
g_{n_h,m_m} \, = \, N_c^{\, (1 - \frac{n_h + n_m}{2})} \;.
\label{coup}
\end{equation}

  This result
may seem a bit surprising to someone steeped in the folklore of large $N_c$ QCD.
It is well known  that in large $N_c$ QCD quark-antiquark components
do not mix with glue states\cite{NC2,NC3}.  It is also well known that for large
$N_c$, exotic mesons do not exist
\cite{NC3}.  These facts are not in conflict with the present analysis.
The lack of mixing between quark-antiquark and glue states concerns
$\overline{q} q$ admixtures into color singlet glueball states \cite{NC2,NC3}. 
In contrast, the exotic quantum number hybrid mesons have color octet glue combined
with a color octet $\overline{q} q$ pair.  The proof of the nonexistence of
exotic mesons applies to $\overline{q} q \overline{q} q$ states only, and not
to states with the quantum numbers of hybrids \cite{NC3}.

The derivation closely parallels the derivation in the purely
mesonic case.   The key ingredient is the calculation of n-point connected correlation
functions of currents which carry the relevant quantum numbers.  For example,
a current for the $\rho$ meson can be written as
\begin{equation}
J_\rho \, = \, \overline{q_i} \, \tau_a \, \gamma_\mu  \, q_i \;,
\label{m}
\end{equation}
where $i$ represents the color index.
An example of a current which creates states with quantum numbers of an
 exotic 
hybrid is the  $I=1$, $J^{pc} = 1^{-+}$ current.  The simplest
current with these quantum numbers is  
\begin{equation}
 J_{h ( 1^{-+})} \, = \, \, g \, \overline{q_i} \,  \gamma_\mu \, \tau_a \, 
F^{\mu \nu}_{i j} \, q_j \;.
\label{h}\end{equation}
It is also useful to consider currents which create glueball states.  An
example which creates the $0^+$ glueball is 
\begin{equation}
J_{g 0^+} \, = \, F_{\mu \nu \,  i j} F^{\mu \nu}_{j i} \;.
\end{equation} 
Generically, the simplest currents with meson quantum numbers are local 
$\overline{q} q$ operators; the simplest hybrids are  $g \overline{q}
F_{\mu \nu} q$ and the simplest glueballs are $F F$.  Here we will
restrict our consideration to these simplest forms; none of the conclusions
depend on this restriction.  To ensure that the hybrid currents do not create 
ordinary mesons one can
restrict attention to those hybrids with exotic quantum numbers.  In what
follows, all of the hybrid sources included will  have exotic quantum numbers.

The n-point functions of the currents may be calculated in the following way:
First  make the 
 replacement,
$ S_{QCD} \, \rightarrow 
S_{QCD} \, - \, \int \, d^4 \,  x  \, \sum_i j_i J_i $, with $j_i$ as classical 
c-number functions and $i$ representing all of the different meson, hybrid and
glueball 
currents of interest. The partition function in the presence of these sources can
be  denoted as $Z[j]$;
the connected n-point functions of interest are given by 
functional derivatives of $\log{Z}$:
\begin{equation}
\langle \, T [J_1(x_1)\,  J_2 (x_2) \,  \ldots \,  J_n(x_n)] \, \rangle_c \,
= \, i^n \, \left . \frac{\partial^n \log (Z[j] )}{\partial j_1(x_1) \,
\partial j_2(x_2) \ldots \partial j_n(x_n)} \right |_{j=0}\;.
\label{mpf}\end{equation}

In the absence of source terms,  standard analysis in terms
of `t Hooft diagrams yields a leading order contribution $\log (Z) \sim N_c^2$
which comes from the sum of all planar gluonic graphs.  The leading order
contributions containing at least one quark line is the sum of all
planar graphs with a single quark loop forming the boundary
of the graph \cite{NC1,NC2,NC3}.  Consider now what happens if the source terms are
included. 

 A typical  mesonic source term will  have  the form 
$J_{m,l} = \overline{q} C_l q$ where $l$ labels the particular mesonic current
and $C_l$ is a matrix in Dirac and flavor space.  The inclusion of such 
terms in the action modifies nothing except the quark propagator.  In
particular one simply makes the following substitution in every diagram:
\begin{eqnarray}
\frac{1}{i \dslash\, - \,m} \,  & \rightarrow &   
\, \frac{1}{i \dslash - m - \sum_l
j_l C_l}\\
 \nonumber 
& =  &\, \frac{1}{i \dslash\, - \,m} \, \,  +  \, \, 
 \frac{1}{i \dslash\, - \,m} \, \sum_l \,  j_l \, C_l 
 \frac{1}{i \dslash\, - \,m} \,  + \, \ldots
\end{eqnarray}
The important point to notice is that this substitution does not affect the
$N_c$ counting at all---$C_l$ is independent of color and hence the
propagators including the mesonic sources carry precisely the same $N_c$ 
factors as those without.

Next consider the effects of including a hybrid source.  The hybrid sources are
all of  the form of quark-gluon vertices.  Thus the inclusion of such
sources simply requires one to replace all quark-gluon vertices by the original
vertex + $\sum_k j_{h,k} \Gamma_k$, where $k$ labels the particular hybrid 
current and $\Gamma_k$ is the quark-gluon vertex associated with it.
The key to the present analysis is that inclusion of these hybrid currents also
does not change the $N_c$ counting.  The color flow through the original QCD 
vertex is the same as that through the vertex induced by the hybrid current---both form two quark legs in the fundamental representation coupled to a
gluon in the adjoint. For convenience a factor of $g$ was included in the
definition of the hybrid current in eq.~(\ref{h}) and other hybrids so that the
$N_c$ factors associated with the coupling constants are not altered by the
substitution.

The inclusion of glueball sources is similar to the previous two cases.
Because of the nonabelian nature of the theory, glueball sources modify both
the gluon propagators and gluon-gluon vertices.   Again, they do not alter the
color flow at all and hence do not change the $N_c$ counting.

Thus, inclusion of  meson, hybrid and glueball source terms does not alter the 
$N_c$ structure of the QCD graphs.  Since all multi-point functions of meson or
hybrid currents involve couplings to at least one quark line while ones
involving the  glueball can couple to pure glue, one
deduces from the $N_c$ scaling of the QCD graphs that 
\begin{equation}
\log ( Z[j_m,j_h,j_g] ) = N_c^2 \, w_2[j_g] + N_C \,  w_1[j_m, j_h, j_g]\;,
\label{lz}\end{equation}
 where $j_m, j_h, j_g$ represent all of the strength of the sources associated
 with mesons,
 hybrids and gluons respectively; $w_2$ and $w_1$ are functionals of the $j$'s.
Note $w_2$ does not depend on $j_m$ or $j_h$.  From eqs.~(\ref{mpf}) and
(\ref{lz}) one sees that connected multi-point functions of $n_h$ hybrid
currents, $n_m$ meson currents, and $n_g$ glueball currents
 have the following $N_c$ dependence provided  there is at least one meson or
 hybrid field:
\begin{equation}
\langle \, T\, [ J_{h 1}(x_{h 1})\, J_{h 2}(x_{h 2}) \, 
\ldots \,J_{h n_h}(x_{h n_h}) 
J_{m 1}(x_{m 1})\,  \ldots \,J_{m n_m}(x_{m n_m}) \, J_{g 1}(x_{g 1}) \, 
\ldots J_{g n_g}(x_{g n_g}) \, ] \rangle_c
\, \sim N_c\;.
\label{mpnc}\end{equation}
If there are no meson or hybrid currents,
\begin{equation}
\langle \, T\, [  J_{g 1}(x_{g 1}) \, T_{g 2}(x_{g_2})
\ldots J_{g n_g}(x_{g n_g}) \, ] \rangle_c\;.
\, \sim N_c^2 \label{gmpnc}
\end{equation}

It is well known in standard large $N_c$ analysis that meson and glueball
interpolating fields create only  single-particle states at leading order. 
 The $N_c$ dependence of
these is \cite{NC2,NC3}
\begin{eqnarray}
\langle m \, | \, J_m \, | \, \rm vac \, \rangle \, & \sim &\, 
N_c^{\frac{1}{2}} \;,\nonumber \\
\langle g \, | \, J_g \, | \, \rm vac \, \rangle \, & \sim &\, N_c \;.
\label{mg} \end{eqnarray}
In an analogous manner one can establish that $J_h$ creates  single-particle 
hybrid states with the following $N_c$ dependence:
\begin{equation}
\langle h \, | \, J_h \, | \, \rm vac \, \rangle \,  \sim \, 
N_c^{\frac{1}{2}} \;.
\label{hnc}\end{equation}
The analyis begins with a consideration of the
 spectral representation for the two-point function for 
currents with exotic hybrid quantum numbers:
\begin{equation}
\int \, d^4 \,x \, e^{-i q \cdot x} \, \langle \, T\,
 [ J^\dagger_{h}(x)\, J_{h}(0) ] \, \rangle \, = 
 \int \, d \,s \, \frac{\rho(s)}{q^2 -s + i \epsilon} \;,
\label{2p}\end{equation}
where the spectral density is $\rho(s)$ and is given by
\begin{equation}
\rho (s) \, = \, \sum_i \, |\langle \, i \, | J_h| {\rm vac} \rangle |^2 \,
\delta(s - M^2) \; ; 
\end{equation}
$| i \rangle$ represents all possible states created by the current.
Since the left-hand side of eq. (\ref{2p}) is of order $N_c$, from eq.~(\ref{mpnc})
one immediately sees that there must exist contributions to $\rho(s)$ which are also of
order $N_c$; hence  there exist  states $i$ for which 
$\langle i| J_h | {\rm vac} \rangle \sim N^{\frac{1}{2}}$.

Before showing that $J_h$ only creates  single-particle states, it is
instructive to demonstrate that consistency of  the spectral decomposition along with eqs.~(\ref{mpnc})
and (\ref{mg}) requires the existence of exotic states.  The method is by
contradiction.  Assuming confinement and no exotics, the only states which
contribute to the spectral function $\rho(s)$ are multi-particle states made of
ordinary meson and glueballs.  In such a case, it follows that there must exist
some class of states with $n_m$ mesons and $m_g$ glueballs for which
\begin{equation} 
\langle n_m \,
  {\rm mesons}; \,  n_g \, {\rm glueballs} \,  | \,  J_h \, |
{\rm vac} \rangle \, \sim \, N_c^{\frac{1}{2}}\;.
\label{s}\end{equation}
Having created this multi-particle state at some time using the curent, $J_h$,  one can 
annihilate it by acting at subsquent times using $n_m$  $J_m$ currents and $m_g$
$J_g$ currents.  From eq.~(\ref{s}) and eq.~(\ref{mg}) one  sees that this
implies
\begin{equation}
\langle \, T\, [ \,J_{m 1}(x_{m 1}) \ldots  J_{m n_m}(x_{m n_m}) \, J_{g 1}(x_{g 1}) \, 
\ldots J_{g n_g}(x_{g n_g}) \, J_h (0)] \rangle_c
\, \sim N_c^{(\frac{1}{2} + \frac{n_m}{2} + n_g)}\;.
\label{cont}
\end{equation}
Equation (\ref {cont}) is in contradiction with eq.~(\ref{mpnc}) unless $n_g =
0$ and $n_m = 1$.  However, $n_g =
0$ and $n_m = 1$ corresponds to a single meson state;
if the quantum numbers are exotic  ({\it e.g.}, $J^{p c} = 1^{- +}$), then
there are no single meson states and one has  a contradiction.
 One concludes that
the assumption that no quantum number exotic states exist is wrong.
 
 To show that $J_h$ only creates  single-particle hybrid states at leading order,
 consider the states contributing to the spectral function.  Recall
$\rho (s)$ is simply $\pi$ times the imaginary part of the two-point function.
Next, consider the imaginary part of the two-point function described diagrammatically
in terms of QCD.  This can be found  by cutting the diagram in all
possible  ways.  Following  Witten's  analysis in ref. \cite{NC2} one 
 sees that to
leading order in $N_c$, all such cuts are composed of quark-antiquark and
multiple gluon states which form a color singlet combination that
cannot be decomposed into more than one color singlet combination.  
Assuming confinement,  this implies that only  single-particle asymptotic 
states are formed to leading order in $N_c$ since two or  more particle
states would require two or more distinct color singlet combinations of quarks
and gluons.  From this, one deduces that at large $N_c$, the
only states $| i \rangle$ which contribute
to the spectral sum are  single-particle exotic hybrid states and that the coupling
strength is given by eq.~(\ref{hnc}).
The two-point function is therefore given by a sum of pole terms:
\begin{equation}
\int \, d^4 \,x \, e^{-i q \cdot x} \, \langle \,
 T\, [ J^\dagger_{h}(x)\, J_{h}(0) ] \, \rangle \, =  \,
 N_c \sum_k \, \frac{|c_k|^2}{q^2 - m_k^2 + i \epsilon}\;.
\label{2p2}\end{equation}

At this stage it becomes clear that so far as $N_c$ counting is concerned
hybrids and mesons behave in exactly the same way.  All of the standard large
$N_c$ analysis of mesonic widths, n-meson coupling strengths, couplings to
glueballs, and so forth, depend only on the fact that multi-point functions involving
at least one meson current are all of order $N_c$, and that $J_m$  ($J_g$) creates
 single-particle  states with an amplitude $\sim N_c^{\frac{1}{2}}$  ($\sim N_c)$
\cite{NC2,NC3}.
On the other hand, from eq.~(\ref{hnc}) one sees that the amplitude for $J_h$
to make a hybrid is also $\sim N_c^{\frac{1}{2}}$; from eq.~(\ref{mpnc}) one
sees that multi-point functions containing hybrids are also of order $N_c$.
Thus, without further computation one can simply apply {\it all} of the
$N_c$ scaling behavior derived for mesons to   hybrids.
  In particular, the
effective coupling between $n_h$ hybrids and $n_m$ mesons is given by
eq.~(\ref{coup}).  

Consider the case of a hybrid decaying into two mesons,
$n_h=1$, $n_m=2$. Equation~(\ref{coup}) shows that  this amplitude is
$N_c^{-\frac{1}{2}}$ implying that the partial width to two-meson decay is 
$N_c^{- 1}$. Similarly, if phase space allows a decay to two hybrids, or a hybrid
and a meson, one sees
that the partial width is also $N_c^{- 1}$.  The partial width for decays including
glueballs or more than two mesons or hybrids is $N_c^{-2}$ or smaller.  One
concludes that at large $N_c$ hybrids decay predominantly into two-particle
states and that their widths are of order $1/N_c$.

In exact analogy with the meson case,  one can deduce the existence of an infinite number of
distinct  hybrid states for any given quantum number as $N_c \rightarrow
\infty$.  The basic strategy is to consider the two-point function 
for large space-like $q^2$ ({\it i.e.}, $q^2 = - Q^2$).  In such a regime, perturbative QCD is valid, and
the two-point function is dominated by diagrams with the fewest number of 
coupling constants.  In this case, this is a two-loop graph.  Using simple
dimensional analysis, it is clear that such  graphs must go as $Q^6$ modulo
logarithmic corrections.  On the other hand, from eq.~(\ref{2p2}) one sees that
if there were only a finite number of hybrid states contributing, the
correlation function at large $Q^2$ would decay as $1/Q^2$.  The only way to
reconcile this is if there are an infinite number of hybrid states with given quantum
numbers.  As a practical matter, for  any given finite $N_c$, the width
of the resonance grows as one goes to higher masses as there is increased phase
space for the decay.  Thus, for any finite $N_c$ there will only be a finite
number of narrow resonance before the states become wide and overlapping.

The preceding argument shows that in large $N_c$ QCD there must be narrow
states with hybrid quantum numbers.  What does this tell us about the real
world of $N_c=3$?  If  the large $N_c$ world is a reasonable
caricature of the physical world the argument suggests that the spectrum should
contain hybrid states with widths comparable to those of mesons.  In a
certain sense, this is not a very strong prediction;  typical mesonic
widths are strongly channel dependent due to both the details of the dynamics
of the coupling and the phase space available for decay.  Therefore it remains
possible that  all of the quantum number exotic hybrid widths are 
``accidentally''  large.
Nevertheless, the analysis presented in this letter is significant: it is
the first argument based on  a systematic expansion (as opposed to a model)
which suggests that narrow exotics states exist in the QCD spectrum.

The author wishes to thank Eric Swanson for raising the question of  how hybrids behave in large $N_c$ QCD,
and thanks Manoj Banerjee and Xiangdong Ji for interesting discussions.
  The financial support of the U.S. Department of Energy under grant number DE-FG02-93ER-40762 
is gratefully acknowledged.

\end{document}